\documentclass[checkin,showpacs,aps,pra]{revtex4-1}
\newcommand{\ba}{\begin{eqnarray}}
\newcommand{\ea}{\end{eqnarray}}

\usepackage{graphicx,color}

\renewcommand{\k}{{\bf k} }
\newcommand{\ke}{{\bf k}}

\newcommand{\be}{\begin{equation}}
\newcommand{\ee}{\end{equation}}
\newcommand{\la}{\langle}

\newcommand{\ra}{\rangle}
\newcommand{\et}{{\it et al. }}
\newcommand{\ete}{{\it et al.}}

\begin{document}

\title{Laser-induced ultrafast transport and demagnetization at the
  earliest time: First-principles and real-time investigation}
\author{G. P. Zhang$^*$} \affiliation{Department of Physics, Indiana
  State University, Terre Haute, Indiana 47809, USA}
\author{Y. H. Bai} \affiliation{Office of Information Technology,
  Indiana State University, Terre Haute, Indiana 47809, USA}
\author{Tyler Jenkins} \affiliation{Department of Physics, Indiana
  State University, Terre Haute, IN.  47809, USA}

\author{Thomas F. George}

\affiliation{Office of the Chancellor
  \\Departments of Chemistry \& Biochemistry and Physics \& Astronomy
  \\University of Missouri-St. Louis, St.  Louis, MO 63121, USA }

\date{\today}

\begin{abstract}
{ It is generally believed that there are at least two
  ways to use an ultrafast laser pulse to demagnetize a magnetic
  sample. One is to directly photo-demagnetize the system through
  spin-orbit coupling (SOC), and the other is to utilize ultrafast hot
  electron transport without SOC. The challenge is that these two
  processes are entangled on the same time scale. While the
  experimental results have been inconclusive, theoretical
  investigations are even scarcer, beyond those earlier studies based
  on spin superdiffusion. For instance, we even do not know how fast
  electrons move under laser excitation and how far they move.  Here
  we carry out a first-principles time-dependent calculation to
  investigate how fast electrons actually move under laser excitation
  and how large the electron transport affects demagnetization on the
  shortest time scale. To take into account the transport effect, we
  implement the intraband transition in our theory.  In the bulk fcc
  Ni, we find the effect of the spin transport on the demagnetization
  is extremely small, no more than 1\%.  The collective electron
  velocity in Ni is 0.4 $\rm \AA/fs$, much smaller than the Fermi
  velocity, and the collective displacement is no more than 0.1 $\rm
  \AA$.  But this does not mean that electrons do not travel fast;
  instead we find that electron velocities at two opposite crystal
  momenta cancel each other. We follow the $\Gamma$-X line and find a
  huge dispersion in the velocities in the crystal momentum space.  In
  the Fe/W(110) thin film, the overall demagnetization is larger than
  Ni, and the Fermi velocity is higher than Ni. However, the effect of
  the spin transport is still small in the Fe/W(110) thin film.  Based
  on our numerical results and existing experimental findings, we
  propose a different mechanism that can explain two
  latest experimental results.  Our finding sheds new light on the
  effect of ballistic transport on demagnetization. }
\end{abstract}

 \maketitle


\section{Introduction}

Interaction between light and magnetism has a long history and can be
traced back to Faraday, Kerr and Voigt effects. Using light to control
and manipulate magnetic properties has become a focus of
research. Over two decades ago, Beaurepaire and coworkers \cite{eric}
discovered that a 60-fs laser pulse could demagnetize a ferromagnetic
nickel thin film within 1 ps. The film was 22 nm thick, coated with a
100-nm MgF$_2$. However, its underlying mechanism is under intense
debate \cite{prl00,koopmans2010,essert2011,baral2014}.  We proposed a
model (H\"ubner model) \cite{prl00,jpcm10} that is based on the direct
interaction of the laser field and spin system via spin-orbit
coupling. The spin-orbit coupling is necessary since it breaks the
spin symmetry and allows the electron to transfer its spin to its
orbital degree of freedom and back \cite{prb08}. This is different
from the magnon picture, where the demagnetization is perceived as the
number of magnon increases as temperature increases, and because the
total spin is still a good quantum number, one has to manually break
the spin symmetry when temperature changes.  In the H\"ubner model,
electrons are itinerant and mobile.  The demagnetization is realized
because the spin expectation value is smaller in conduction bands than
valence bands close to the Fermi surface. Koopmans \et
\cite{koopmans2010} proposed a similar model by emphasizing on spin
mixing and spin flipping through the phonon interaction. Naturally,
spin mixing and spin flipping are also included in the H\"ubner
model. The key difference between the Koopmans model and H\"ubner
model is the way that the spins move out of the system.

Battiao \et proposed a different model, the spin superdiffusion model
(SSD) \cite{battiato2010,battiato2012}.  SSD does not need the
spin-orbit coupling, but relies on the difference between majority
spin and minority spin diffusions. Since majority spins move faster
than minority spins, this creates a depletion zone for majority
spins. Assuming the minority spins stay, loss of the majority spins in
the excited regions leads to demagnetization.  They argued that SSD
can even completely explain the ultrafast demagnetization
\cite{battiato2010}.  Melnikov \et \cite{melnikov2011} carried out the
second-harmonic generation measurement and found that upon  laser
pumping on the Fe layer, the gold layer becomes spin polarized with a
clear hysteresis loop.  However, this experiment only showed the spin
transport out of Fe to Au layer, and did not prove that SSD is
responsible for demagnetization.  Vodungbo \et \cite{vodungbo2012}
examined a faster demagnetization (within 100 fs) in CoPd multilayers,
with each stack as  1 nm thick. They found no modification of the
magnetic structure and the resonant magnetic scattering patterns peaks
at the same wave vector transfer. Nevertheless, they assigned this
finding to the direct transfer of spin angular momentum between
neighboring domains.  About one month later, Pfau \et \cite{pfau2012}
carried out a similar experiment in CoPt multilayers and reached a
different conclusion that the peak of the small-angle x-ray scattering
shifts with time.

Two days later after Pfau's submission, Rudolf \et \cite{rudolf2012}
reported that the ultrafast magnetization enhancement was driven by a
superdiffusive spin current. They found that in the Ni/Ru/Fe trilayers
if the magnetizations in the Ni and Fe layers are parallel to each
other, the magnetization in the Fe layer increases.  However, the
amount of decrease in the Ni layer is not equal to the amount of
increase in the Fe layer. In addition, they found that there is a
laser fluence limit of 2.0 mJ/$\rm cm^2$, beyond which only the
demagnetization is observed. They argued that this was due to the spin
superdiffusion saturation.

Eschenlohr \et \cite{eschenlohr2013} identified the ultrafast spin
transport as the sole mechanism for femtosecond demagnetization,
excluding spin-flips that are directly induced through the spin-laser
field interaction \cite{prl00,np09,bigot09}. They showed that SSD
could accurately explain their observation.  Experimentally, they
employed x-ray circular dichroism to probe the spin change in an Au/Ni
layered structure. They shined the light directly on to the thicker
nonmagnetic Au layer, so the direct light excitation of Ni is
smaller. In this case, only hot electrons hit the nickel layer.  They
concluded that the fact that the Ni layer is demagnetized shows the
transport is the dominant factor, which excludes other mechanisms
such as spin flip or spin-laser interaction. Thus, they believed that
they provided decisive arguments for unraveling the origin of
ultrafast demagnetization.  In the same year, Turgut \et
 \cite{turgut2013} showed that in contrast to the earlier findings
 \cite{battiato2010,battiato2012,eschenlohr2013}, the spin
superdiffusion is not the only process that leads to the
demagnetization. They found that if they reversed the order of the
multilayer by placing the iron layer first and then the nickel layer,
there is no spin enhancement. This shows that the appearance of spin
diffusion is system dependent, not intrinsic to the demagnetization.

A more direct test is from the work by Schellenkens \et
\cite{schellekens2013}. They grew wedged Ni films on both insulating
sapphire and conducting aluminum substrates, exactly the same as those
used in the theory \cite{battiato2010,battiato2012}. But to their
surprise, the temporal evolution of the magnetization, regardless of
whether it is pumped on the front or on the back of the sample, is
identical. They argued that if back pumped, the spin should accumulate
in the front, and when the probe pulse detects it, the outgoing signal
should be increased. They also purposely reduced the laser intensity
so SSD can play a role, but no influence of transport was
found. However, the same group \cite{hofherr2017} reached a different
conclusion lately for the Ni/Au system, where they had substantial
evidence that the demagnetization was dominantly driven by spin
currents.  Moisan \et \cite{moisan2014} showed that regardless of
their sample magnetic configuration, the demagnetization time remains
the same, and they concluded that hot electron spin transfer between
neighboring domains does not change the ultrafast
magnetization. However, they suggested that the effect of spin
transport on demagnetization may be related to the spin accumulation
length.  von Korff Schmising \et \cite{vonkorff2014} attempted to
image the demagnetization dynamics using a holographic mask.  They
found a rapid lateral increase of the demagnetized area, with the
propagation front moving with a speed on the order of 0.2
nm/fs. However it is difficult to correlate the demagnetization with
the lateral increase.

Shokeen \et \cite{shokeen2017} employed a 10-fs pulse to probe the
magnetization dynamics in Ni and Co systems of various thickness from
10 to 40 nm, and found that ultrafast demagnetization is again system
dependent, and both spin majority and minority channels contribute,
not that the majority alone contributes as assumed in the spin
superdiffusion theory \cite{battiato2010,battiato2012}. An increase in
Co was observed but on a time scale of 20 fs, far shorter than
appropriate for SSD.  {In Co/Cu(001), Chen \et
  \cite{chen2018} further showed that demagnetization does not occur
  through redistribution of spin among Co and Cu atoms, though their
  TDDFT calculation is still unable to reproduce the same amount of
  spin moment reduction as their experiment.}  Tengdin \et
\cite{tengdin2018} showed that demagnetization and the collapse of the
exchange splitting in Ni are mediated by low-energy magnon, not
SSD. However, magnon excitation permutes with the total spin, so it is
puzzling why demagnetization could occur.  In contrast to prior
experimental results \cite{koopmans2010}, they found that the
demagnetization time is fluence independent and is 176 fs.  The origin
of the above experimental discrepancy is unknown.  A quasi-phase
transition at 20 fs is attributed to both SSD and spin-orbit
coupling. But on such a short time scale, transport on the 1 nm scale
should be ballistic, not diffusive, while the spin-orbit coupling
$\lambda$ is too weak (20 fs corresponds to 0.205 eV, and in nickel
$\lambda=0.07$ eV \cite{prl00}).  In CoPt multilayers, Zhang \et
\cite{zhang2018} showed that the demagnetization is always at 150 fs,
independent of external magnetic field amplitudes. If the spin
transport between different magnetic domains were important to
demagnetization, one would expect that the domain structure must
affect the demagnetization. Their results show this is not the
case. They attributed the local dissipation of spin angular momentum
as a dominant channel to demagnetization. The material specific nature
of demagnetization also appears in NiPd magnetic alloys.  This is an
ideal model system for SSD where Ni and Pd atoms are next to each
other, so the expected spin superdiffusion should be very
strong. However, Gang \et \cite{gang2018} concluded that the optically
triggered spin current between the subsystems of Ni$_x$Pd$_{1-x}$
alloys does not dominate the demagnetization, in contrast to SSD
\cite{battiato2010,battiato2012}. On the other hand, Fert$\acute{\rm
  e}$ \et \cite{ferte2017} showed that the hot-electron pulse can
demagnetize CoTb alloys as well.


So far, there has been no consensus experimentally.  A theoretical
investigation at the initial stage of laser-induced demagnetization
and transport is imperative. This would potentially allow one to
extract useful insights from SSD and develop a new picture.  In this
paper, we employ the first-principles time-dependent Liouville density
functional theory \cite{jpcm16}, without resorting to the empirical
procedure \cite{battiato2010,battiato2012}. We take into account both
the interband transition and intraband transitions (transport effect)
among band states. We find that the effect of direct laser-induced
transport on the demagnetization is very weak. In fcc Ni, the electron
oscillates with a maximum collective velocity amplitude of 0.4
$\rm\AA/fs$, far below the Fermi velocity, and a net displacement of
0.07 $\rm \AA$ within 300 fs. A similar situation is found for one
monolayer Fe on three layers of tungsten. The net spin percentage
change due to the intraband (transport) contribution is only 0.1\%. We
find that although the crystal-momentum dispersed velocities are
large, the strong cancellation of the velocities at two opposite
crystal momentum points results in a small net velocity. Based on our
numerical results and prior experimental findings \cite{bergeard2016},
we propose a new picture to identify the pure transport-induced
demagnetization through the ballistic transport \cite{bergeard2016},
where both majority and minority spins travel at their respective
velocities. This picture allows us to explain two latest experimental
results \cite{shokeen2017,hofherr2017}, without invoking SSD.  Our
study reveals crucial insights into the effect of the transport on the
laser-induced demagnetization.

The rest of the paper is arranged as follows. In Sec. II, we present
our theoretical formalism with details on the intraband transition. We
show our results in Sec. III, where we examine the Fermi velocity and
the velocity change under the laser excitation, followed by the spin
moment change with and without intraband transitions. Section IV is
devoted to the discussion. Finally, we conclude this paper in Sec. V.

\section{Theoretical formalism}

\newcommand{\ik}{n{\bf k}}
\newcommand{\jk}{m{\bf k}}

In traditional spin transport, an external bias is applied
longitudinally along a sample. Figure \ref{fig1}(a) illustrates such
an example, where the electric field points to the left and the
electrons move to the right. This is very different from laser-induced
spin transport (see Fig. \ref{fig1}(b)). Light is a transverse wave,
where its electric field ($x$ axis) is perpendicular to the laser
propagation direction ($z$ axis). Therefore, initially electrons must
move along the $x$ axis, not along the $z$ axis as assumed in several
previous studies
\cite{eschenlohr2013,seifert2018,li2018}. {We note in
  passing that all the velocities here refer to the instantaneous
  velocities, not the time-averaged one.} Only after this initial
interaction with the laser field may the electrons that are close to
the surface of a sample scatter with electrons that are away from the
surface. It is this initial interaction of the electrons with the
laser field that initiates laser-induced spin dynamics and spin
transport, and underlies all the steps of laser-induced ultrafast
demagnetization \cite{eric}, a hot topic that remains unsolved up to
now \cite{ourreview,rasingreview}.

Our theory starts with the standard density functional theory as
implemented in the Wien2k code \cite{wien2k}. We first solve the
Kohn-Sham equation (in atomic units) \cite{prb09}, \be
[-\nabla^2+V_{ne}+V_{ee}+V_{xc}]\psi_{\ik}(r)=E_{\ik} \psi_{\ik}
(r), \label{ks} \ee to find the eigenvalues and eigenvectors. The
terms on the left-hand side represent the kinetic energy,
nuclear-electron attraction, electron-electron Coulomb repulsion and
exchange correlation, respectively. We use the generalized gradient
approximation (GGA) at the PBE level \cite{pbe}.  $\psi_{\ik}(r)$
represents the Bloch wavefunction of band $n$ at crystal momentum
${\bf k}$, and $E_{\ik}$ is its band energy. These wavefunctions are
used to construct the optical transition matrices for the
time-dependent calculations. In the original Wien2k code, the matrix
elements ($-i\nabla$ operator) are stored with a precision to
$10^{-6}$. We modify the code so we can store the entire matrices
unformatted, thus keeping all the 16 significant figures.  The
spin-orbit coupling (SOC) is included using a second-variational
method, where spin-polarized eigenstates are used as the basis for the
SOC calculation. The spin-matrix is constructed among band states by
our home-built code that obeys the regular spin permutations
\cite{prb09}.

To investigate the spin transport, we construct the electron velocity
operator from the momentum operator as $\hat{v}=-i\hbar \nabla /m_e$,
where $m_e$ is the electron mass. In the absence of an external field,
electrons on the Fermi surface travel with the Fermi velocity $v_f$,
but their net velocity is zero because a nonzero velocity at a ${\bf
  k}$ point cancels another velocity at a $-{\bf k}$ point.  There are
several methods that we can use to compute the Fermi velocity. One is
to take the derivative of the band energy $E_{\ik}$ with respect to
\ke. However, this may run into a singularity issue if the band
dispersion is too steep, so we use a different method.  After the
convergence of our self-consistent calculation, we compute the
momentum matrix elements between band states at each ${\bf k}$ point,
\be \la \ik |\hat{\bf P} |\jk\ra= \la \psi_{\ik} |-i\hbar \nabla |
\psi_{\jk} \ra, \ee where $\psi_{\ik}(r)$ and $\psi_{\jk}(r)$ are the
wavefunctions for the band states $\ik$ and $\jk$, respectively.  The
diagonal matrix element of $\la \ik |\hat{\bf P} |\ik\ra$ is used to
find the velocity $v_{\ik}=|\la \ik |\hat{\bf P} |\ik \ra/m_e|$, where
$m_e$ is the electron mass.  To compute the Fermi velocity, we
integrate $v_{\ik}$ over \k and sum over all those states on the Fermi
surface, \be v_f=\sum_n\int d{\bf k} v_{\ik} \delta
(E_{\ik}-E_f), \label{fermi} \ee where $E_f$ is the Fermi energy and
$E_{\ik}$ is the band energy.  The $\delta$ function is replaced by a
broadening $\epsilon$ in the actual calculation, such that the states
with energy $|(E_{nk}-E_f)|\le \epsilon$ are included in the
integration.

Our real time-dependent simulation starts with the Liouville equation
for density matrices $\{\rho_\k \}$ at every \k point
\cite{prb09,jpcm16,dignam2014,luu2016}, \be i\hbar \frac{\partial
  \rho_\k}{\partial t} =[H_0+H_I, \rho_\k] -ie {\bf F}(t)\cdot
\nabla_\k \rho_\k \label{liu} \ee where $H_0$ is the field-free system
Hamiltonian.  The interaction between the laser and system is $H_I=-e
{\bf F}(t) \cdot \sum_\k{\bf D}_\k\rho_\k$, where ${\bf F}(t)$ is the
laser electric field with the amplitude $F_0$ in $\rm V/\AA$ and has a
Gaussian shape with pulse duration $\tau$ in fs. The laser photon
energy is $\hbar\omega$.  The normal Liouville equation \cite{prb09}
is recovered if the second term on the right side of Eq. (\ref{liu})
is absent.  This second term is the intraband transition term between
different \k points and is directly responsible for electron transport
between different \k points. However, this introduces a numerical
complication that the density matrices at different \k points are no
longer separable, and numerical calculations become very time
consuming since the \k parallelization is not possible.  A technical
detail should be mentioned here \cite{dignam2014,luu2016}. In
Eq. (\ref{liu}), the second term on the right side should be treated
with great care \cite{luu2016}. We use the fourth-order derivative
solver and use a dense \k mesh grid, which guarantees the accuracy of
our calculation. {In the case that interband transitions are ignored,
  the effect of the laser field is equivalent to shifting the Fermi
  sphere as shown in Fig. \ref{fig1} (for details, see the
  Appendix). In our calculation, we directly use Eq. (\ref{liu}), so
  both intraband and interband transitions are included.} {
  Our method is similar to the time-dependent density functional
  theory\cite{krieger2015,elliott2016,krieger2017,dewhurst2018}, and
  rigorously obeys the Pauli exclusion principle, so we can
  investigate the electron population change dynamically. We use the
  length gauge since it allows us to separate the intraband and
  interband transitions easily because they appear in two separate
  terms in our Liouville equation. For this reason, the length gauge
  has been frequently used for solids \cite{golde2008,dignam2014,luu2016}.
}

\section{Results}

Before a laser field interacts with a system, electrons on the Fermi
surface travel with the Fermi velocity. The laser field exerts an
additional force on those electrons. Most of prior studies do not
address some of the basic questions in transport.  For instance, how
fast do the itinerant electrons move under laser excitation? How far
do they transport? In regular diffusion processes, there must be a
gradient between different parts of a sample.  Our goal is to develop
a picture for electron transport on a solid ground and investigate how
much the laser impacts the electron dynamics on the shortest time
scale. We consider two systems, one bulk and one thin film. We choose
bulk fcc Ni and a thin film with one monolayer of iron on top of three
layers of tungsten in a slab geometry.  We can not think of a better
place to start with transport by looking at the Fermi velocity.

\subsection{Fermi velocity in Ni}

We start with fcc Ni. In our calculation, we adopt a simple cubic
structure (4 Ni atoms per unit cell) to avoid the issue of the
derivative of the density matrix with respect to the crystal
momentum. We use the \k points in the full Brillouin zone instead of
the irreducible one for the same reason. The size of our problem is
determined by the number of \k points $N_k$ and the number of bands
$N_b$. The matrix size is $N_kN_b\times N_kN_b$. Given the limit of
our computer resource, we can only adopt a \k mesh of $16\times 16
\time 16 \times 16$ and $N_b=60$. We remove 32 low-lying states (8
states, 2 for $3s$ and 6 for $3p$ per Ni atom), so these 60 states
span across the Fermi level and reach all the way up to 1 Rydberg,
which is more than enough to cover all the bands affected by the laser
excitation.

In the discrete mesh, the Fermi surface is not clear cut. We have to
use a broadening in the form of a shell around it. This broadening has
a physical meaning as well if we consider it as a thermal broadening
that can be changed.  We use Eq. (\ref{fermi}) to compute the Fermi
velocity. Figure \ref{vf} shows our theoretical Fermi velocity in fcc
Ni as a function of the broadening $\epsilon$ around the Fermi energy
$E_f$.  $\epsilon$ allows us to control the number of band states
entering the integration in Eq. (\ref{fermi}).  We see that the Fermi
velocity has a nontrivial dependence on $\epsilon$, but in general it
decreases with $\epsilon$.  The vertical dashed line denotes the room
temperature broadening. The crossing point on the curve gives us our
theoretical velocity $v_f=2.79~\rm \AA/fs$, which is in an excellent
agreement with the experimental value of 2.8 $\rm \AA/fs$ by Petrovykh
\et \cite{petrovykh1998} (the horizontal dotted line in
Fig. \ref{vf}(a)). This demonstrate the high accuracy of our
calculation.

Although the electrons around the Fermi surface move with $v_f$, there
is no net current or transport. This is because for every velocity at
${\bf k}$ point, there is a velocity in the opposite direction $-{\bf
  k}$ point. Physically, electrons at $\pm {\bf k}$ move in opposite
directions, so the net current is balanced out. Figure \ref{vf} shows
one example of velocities for the energy band $n=70$ at ${\bf
  k}_1=[(11,15,11)/32] b$ and ${\bf k}_2=[(-11,-15,-11)/32] b$, where
$b$ is the reciprocal lattice vector.  We see indeed \be {\bf v}({\bf
  k}_1)=-{\bf v}({\bf k}_2). \label{cancel} \ee All three components
are numerically exactly the same. Therefore, when one discusses how
fast electrons move, one must consider electrons at both ${\bf k}$ and
$-{\bf k}$ points. The net spin change carried by those two electrons
must be summed up to zero in the absence of an external field. The
actual velocity that one should use for spin transport is not $v_f$,
but the net velocity is $v_{\rm net}=v_{\rm laser}-v_{\rm without~
  laser}$. This is because $v_{\rm without~ laser}$ allows electrons
to reach the thermal equilibrium, while the extra velocity due to the
laser field allows electrons to move out of equilibrium. In the next
subsection, we compute how fast the electrons move collectively.

\subsection{Velocity change under laser field excitation}

Central to transport is the electron motion.  It is interesting to
note that there has been no study based on SSD to directly compute the
electron velocity. We fill this important gap.  We choose a linearly
$x$-polarized pulse of $\tau=60$ fs, $F_0=0.03\rm ~V/\AA$ and
$\hbar\omega=2$ eV, propagating along the $z$ axis. Our laser field
amplitude is comparable to experimental values \cite{koopmans2010},
and at the field maximum, this corresponds to the crystal momentum
shift $\Delta k=0.015/\rm \AA$. Since the reciprocal lattice vector
length in fcc Ni is $b=2\pi/a=2\pi/3.51882=1.7856\rm/ \AA$, $\Delta k$
represents only 8.4/1000 of the Brillouin zone, extremely small.
Light is a transverse wave, and its electric field must be
perpendicular to the propagation direction. If the light propagates
along the $z$ axis, electrons experience no external force along the
$z$ axis initially.  This observation has apparently evaded prior
investigations
\cite{battiato2012,eschenlohr2013,rudolf2012,bergeard2016}.

Our numerical result confirms the above observation. Figure
\ref{vf}(c) shows the system averaged velocity along the $x$ axis,
$v_x=\sum_k{\rm Tr} (\hat{v}^x_\k\rho_\k)$, as a function of
time. Velocities along the other directions are much smaller.  Our
laser pulse peaks at 0 fs. From the figure, we see that $v_x$
increases sharply, already starting at -100 fs, and peaks at -20 fs,
ahead of the laser peak.  $v_x$ oscillates rapidly between $-0.4\rm
~\AA/fs$ and $0.4\rm~ \AA/fs$.  This velocity is only 14\% the Fermi
velocity.

The key premise of SSD is that laser-excited electrons in $sp$ bands
are transported and $d$ electrons are treated as local
\cite{battiato2010}.  The theory is based on a prior static
calculation \cite{zhukov2006} where the $sp$ electrons have a speed of
10 $\rm\AA/fs$.  The argument is that if one puts electrons in states
2 eV (photon energy) above the Fermi level, they acquire this
velocity. To be sure, we also calculate the same static
crystal-momentum averaged velocity as a function of the energy
referenced to the Fermi energy. The inset in Fig. \ref{vf}(c) shows
that electrons at 2 eV can indeed gain 10 $\rm\AA/fs$, consistent with
Zhukov's finding \cite{zhukov2006}, but whether all those $d$
electrons can be excited to 2 eV has been unknown dynamically.

Our calculation gives an answer to this question. We find a much lower
velocity, where the reason is very simple.  In the laser excitation,
there are lots more intermediate states occupied below 2 eV, and
electrons in those states have a lower velocity.  The static
estimation overestimates the level of excitation. Even if the $sp$
electrons move with such a high velocity, their contribution to spin
change would be limited because $sp$ electrons are not strongly spin
polarized and have a very small effect on the demagnetization.  This
10 $\rm\AA/fs$ is 3.5 times larger than the Fermi velocity and 25
times larger than our calculated peak velocity. Furthermore, the
velocity only peaks within a narrow time window, after which it
subsides quickly. For our current laser parameter, this window is
about 50 fs.

The velocity is not the only one that we can examine.  To see whether
electrons indeed diffuse away from their original location, we
integrate the velocity $v_x$ to get the collective displacement of the
electrons, \be \Delta x(t)=\int_{-\infty}^t v_x(t')dt', \ee where
$v_x(t')$ is the velocity along the $x$ axis at time
$t'$. {Note that even though the velocity appears to be
  symmetric, if we zoom in, we find that there is an asymmetry in the
  velocity. This velocity drift accumulates as time evolves and leads
  to the net displacement.}  Figure \ref{vf}(d) shows the displacement
as a function of time. It is clear that the rapid oscillations of the
electrons do not lead to a large net displacement in the position
space. At the end of the pulse, the net displacement is less than 0.1
$\rm \AA$.  In our simulation, we use a simple cubic structure (a
supercell with four Ni atoms) to simulate a fcc structure, so we can
investigate whether electrons transport from one lattice site to
another. Ni's lattice constant is 3.52 $\rm \AA$, so the net transport
effect is very small, which is consistent with our expectation.
However, this does not mean that the electron transport does not
occur, but it means that the laser-induced one is very small at the
earliest stage. This is the time scale that SSD claims to be able to
completely explain the demagnetization \cite{battiato2010}. An
additional challenge for SSD is the direction of the forces that
electrons experience. Without laser excitation, the net force on the
electrons has to be zero. As briefly discussed above, if a laser pulse
propagates along the $z$ axis, the laser electric field must be in the
$xy$ plane. For a tetragonal structure (with the spin-orbit coupling
and magnetic quantization axes along the $z$ axis, a fcc structure
becomes tetragonal), the net force along the $z$ axis is zero by the
space symmetry, at least in the beginning of laser excitation.  This
questions the rationale that SSD always assumes the electron
propagation direction to be along the light propagation direction.


\subsection{Effect of electron transport on demagnetization in bulk nickel}

So far, we have only investigated the electron dynamics, in
particular, how the electron changes its velocity upon laser
excitation.  Next, we see how electron transport affects spin
dynamics. We start from fcc Ni.  The results are shown in
Fig. \ref{ni}. The solid line is the spin moment with the intraband
term in Eq. (\ref{liu}), while the dashed line is without the intraband
term. Figure \ref{ni}(a) shows that both cases have a similar spin
change, and their difference is very small mainly after the
minimum. The recovered spin moment for the non-intraband transition is
larger, i.e., smaller demagnetization.  To see the detailed change, in
Figure \ref{ni}(b) we plot their difference $\Delta M_z=M_z^{\rm
  intra}-M_z^{\rm no ~intra}$ as a function of time. The direct impact
of transport is small, only about 3\%.  It is clear that the intraband
contribution is mainly on a time scale longer than 100 fs, after the
demagnetization maximum.

{We further examine how the velocity disperses with the
  crystal momentum under the laser excitation. This information is
  crucial since it provides the details of electron dynamics.  There
  are many crystal momentum directions that we can examine. We choose
  the $\Gamma$-X direction, since along this direction the laser field
  is applied.  Figure \ref{ni}(c) shows the first half of the
  $\Gamma$-X line, with the crystal momenta value given in the caption
  and denoted in the figure by $k_i$.  Note that our $k$ mesh is
  shifted for convergence purposes.  $k_1$ approximately corresponds
  to the $\Gamma$ point. We see that as we move away from the $\Gamma$
  point, the magnitude of the equilibrium velocities (the base lines)
  is higher as expected. But it only increases up to $k_5$, after
  which the velocity starts to decrease, since the band starts to
  change. It is clear that at each \k point, the electron velocity
  gain differs. We see that at $k_6$, $k_7$, and $k_8$, there is
  little gain, but the gain is large at $k_4$. This is directly
  connected to the band structure itself. So far, all the velocities
  are negative. If we examine the second half of the $\Gamma$-X line,
  we see that those velocities are all positive (see
  Fig. \ref{ni}(d)). This is because the band dispersion changes its
  slope \cite{prb09}. Now if we compare Figs. \ref{ni}(c) and (d),
  these velocities are nearly opposite to each other. In other words,
  in a bulk material, the electrons move in the opposite
  directions. To have a net flow of electrons, the system must have an
  asymmetry. }

\subsection{Effect of electron transport on demagnetization in an
  Fe/W(110) ultrathin film}

In the following, we investigate an ultrathin film, where we place a
monolayer of Fe on the top of three layers of W(110) (see
Fig. \ref{fig1}(d)). To maintain the inversion symmetry, it is
customary that another layer of Fe is placed at the bottom of W. We
adopt a supercell structure where we have added a vacuum layer to
separate these slabs. The thickness of the vacuum layer is 11.19 $\rm
\AA$, or five layers.  We first optimize the structure along the $z$
direction, assuming pseudomorphic growth. The optimized structure has
the Fe atom shifted about 1\% toward the W atom.  The spin moment is
mainly on the Fe atom, 2.5 $\mu_B$, while the tungsten atom has a very
small value of $-0.1\mu_B$. From the above study, we already see the
small change in Ni due to the intraband transition, so we wonder
whether there is any difference in the dipole moment (which reflects
the optical response). Figure \ref{fig2}(a) compares two dipole
moments, with and without the intraband transition. We shift the one
with the intraband transition vertically by one unit for clarity. We
see that there is no visual difference. Figure \ref{fig2}(b) shows
their numerical difference, where we multiply the curve by a factor of
100. We see that the impact of intraband transitions on the dipole
moment is more pronounced. The difference starts earlier before the
laser pulse peaks. This is expected since the dipole reflects charge
response as it responds faster than the spin \cite{jmmm98}.

{The spin moment is plotted as a function of time in
  Fig. \ref{fig2}(c). The solid line (black) is the one without
  intraband transitions, while the dotted line (red) is the one with
  intraband transitions. We see that they almost overlap with each
  other. To see their difference, we multiply it by 1000 and show it
  in Fig. \ref{fig2}(d).  We find the same conclusion is true for an
  Fe/W(110) thin film. The spin change due to the intraband transition
  is very small. However, we see the overall demagnetization is larger
  in the Fe/W(110) ultrathin film than that in Ni (compare
  Figs. \ref{ni}(a) and \ref{fig2}(c)). We wonder whether this is
  connected with the Fermi velocity. Figure \ref{w3fe2}(a) is the
  Fermi velocity as a function of the broadening $\epsilon$. The Fermi
  velocity at room temperature is highlighted with a vertical line. It
  is 3.06 $\rm \AA/fs$, which is indeed higher than that in Ni. Next,
  we also compute the velocity as a function of energy with respect to
  the Fermi energy. If the electrons are all excited to a particular
  energy, they will acquire this velocity.  Figure \ref{w3fe2}(b)
  shows that at 2 eV, the velocity is less than 3 $\rm \AA/fs$, less
  than that at the same energy in Ni. This demonstrates that if the
  velocity at the high energy window is crucial to the
  demagnetization, then we should expect a larger demagnetization in
  Ni. Our data do not support such a scenario.  }

\section{Discussions: Necessity of ultrshort pulses}

In retrospect, many earlier claims have been overstated, without
leaving sufficient room for new ideas.  When we examine the SSD theory
closely, we notice in the initial step how the $sp$ electrons are
excited by a laser pulse is missing. Instead, the entire generation
process is controlled by a source term $S^{\rm ext}$ which is not
given in their publications \cite{battiato2010,battiato2012}. This
prevents one from examining their theory further. However, it becomes
clear now that they made an important assumption that each Ni atom
takes 0.1 photon (with photon energy of 1.5 eV) and each Fe atom takes
1 photon in their theory \cite{battiato2012}. As we showed recently,
this 0.1 photon is sufficient to reproduce all the demagnetization
process in Fe, Co and Ni \cite{jpcm16}, even without invoking spin
superdiffusion. In the H\"ubner model \cite{prl00}, the laser
excitation enters through the dipole interaction term.  {
  The conservation of angular momentum is achieved through the dipole
  transitions, where the laser field and the magnetic system exchange
  orbital angular momentum. The linear momentum of the photons at our
  wavelength is extremely small, in comparison with the electron
  momentum, and is ignored here.}  A similar approach was employed in
the time-dependent density functional theory calculation
\cite{shokeen2017}. This is the standard method that one can
systematically increase the laser amplitude as we did before
\cite{jap11}.  Both the theory \cite{jap11} and experiment
\cite{fognini2015} showed that a shorter laser pulse induces a much
steeper demagnetization, which is significantly different from those
with a longer laser pulse where a more gradual decrease in
magnetization is observed.  A similar laser-fluence dependence in SSD
is unknown.

To understand the role of transport in the demagnetization, we face
multiple challenges. First, both the spin-orbit coupling induced
demagnetization \cite{prl00} and the spin superdiffusion-induced
demagnetization \cite{battiato2010} occur on a similar time scale, so
it is difficult to separate them in the time domain.  Second, there is
a difference between (a) using hot electron transport to demagnetize a
sample and (b) proving that the demagnetization exclusively comes from
hot electron transport. (a) is similar to transient electron
doping. Nickel and copper differ by one valence electron, but one is
magnetic and the other is not. There is no surprise here.  (b) is more
tricky since there are many possible ways that a magnet can be
demagnetized. To demonstrate that demagnetization comes from electron
transport requires an exhaustive effort to exclude all the possible
channels.  Vodungbo \et \cite{vodungbo2016} stated clearly that even
though indirect excitation can lead to ultrafast demagnetization, this
can not be used as evidence for SSD, since the amount of gain and loss
in spin polarization must both be measured to quantitatively determine
the relevance/contribution of superdiffusive spin transport to the
overall demagnetization.  Since demagnetization and spin transport
occur on a similar time scale, it is necessary to employ a shorter
pulse to disentangle their difference.

Next, we outline what should happen if the demagnetization is due to
the ballistic transport alone, given that most of samples are very
thin.  Figure \ref{fig3}(a) shows a case for the ballistic transport
with a short magnetic sample with length $l_m$. We assume that the
laser pumps on the front (the right side) and the detection can be
either in the front or the back.  The times for the majority and
minority spins to travel through the sample are \be
t_\uparrow=l_m/v_\uparrow {~~~~~\rm and~~~~~~~~~}
t_\downarrow=l_m/v_\downarrow,\ee respectively. We take the
experimental parameters from Shokeen \et \cite{shokeen2017}. The
thickness of their film is 10 nm. By using the velocities for the
majority and minority spins \cite{zhukov2006}, the time time delay
$\Delta t_{sp}$ of the minority $sp$ spin at 1.5 eV with respect to
the majority $sp$ spin is 2.6 fs. Therefore, from 0 to 10.5 fs
($t_\uparrow(sp)$), the back side of the sample should show the spin
moment enhancement. After 2.6 fs, the minority spins arrive and the
enhancement stops, so the spin moment returns back to the pre-pump
value. In the meantime, the front probe should see the
demagnetization. If the pump is strong, the magnetic moment should
drop to zero and reverse the sign, since the minority becomes the
majority as the true majority spin moves out of the region. This 2.6
fs is way too short for many experiments to detect $sp$ spin
transport.  However, if the transport is carried by the $3d$ electron
spins, which is not included in the original SSD theory
\cite{battiato2010}, then $\Delta t_{3d}$ is 54.1 fs. This time delay
is within the regime of the experiment \cite{shokeen2017}. The 42-fs
spin enhancement peak observed in the gold layer by Hofherr \et
\cite{hofherr2017}, which is very close to our time of 54.1 fs, is now
explainable, since incidentally their nickel thin film thickness is
exactly the same as that of Shokeen \et \cite{shokeen2017}.  It is
more likely that both majority and minority spins reach the gold
layer. We will come back to this below.

In Fig. \ref{fig3}(b), we schematically show the magnetization change
as a function of time for the front probe and back probe. The ideal
experimental detection is on the back side. The front side probe
suffers from the charge depletion as majority and minority spins move
out of the regime. If an insulator is attached to the front, this
creates a capacitor effect that pulls both majority and minority spins
back, so the magnetic moment crosses zero again. If a conductor is
attached to the front, the electron flow from the conductor to the
ferromagnetic sample further complicates the entire process. On the
other hand, the back side probe is relatively cleaner because charge
carriers tend to move out of the sample. It must show a hump at
$\Delta t$ if the demagnetization is dominated by the spin
transport. From the experimental data \cite{shokeen2017}, if the $sp$
spin transport is important, the peak location is beyond the current
laser pulse duration; if the $3d$ spin transport is important, this
should be detectable, but this was not observed experimentally in Ni
\cite{schellekens2013,shokeen2017}.

For Co, we do not have a good experimental velocity.  Sant \et
\cite{sant2017} estimated the spin diffusion coefficients at 500 fs,
far beyond the superdiffusion limit. Although they implied the results
are from the domain wall, it is more likely that they detected the
spatial spin distribution, rather than the domain wall motion, since
the domain wall can not move so fast. They estimated the spin
diffusion coefficient $D$(at 500 fs) to be 0.35 $\rm nm^2/fs$ for spin
up and 0.02 $\rm nm^2/fs$. We can compute the spin velocity through
\be v_{\uparrow(\downarrow)}=\sqrt{D_{\uparrow(\downarrow)}/t(500~\rm
  fs)}, \label{diffusion} \ee which gives $v_{\uparrow}=0.26 \rm
\AA/fs$ and $v_{\downarrow}=0.063 \rm \AA/fs$.  These velocities are
in line with our theoretical findings (see Fig. \ref{vf}(c)), though
we have a different system and their velocities already pass their
maxima. For the same thickness of 10 nm, if the majority and minority
spins moved with these velocities to traverse the entire sample, the
time delay $\Delta t$ between the spin up and spin down would be 1209
fs.  Next, we extrapolate their diffusion coefficient all the way to 0
ps by a quadratic function, and applying the same equation
(Eq. (\ref{diffusion})), we find the time delay is reduced to 753.9
fs.  This surely over-estimates the delay, but it does point out that
the delay in Co is qualitatively longer than that in Ni.  If we use
the theoretical estimate for the majority spin $v_\uparrow=2.55~\rm
\AA/fs$ \cite{gall2016}, we can figure out the velocity for the
minority spin.  Shokeen \et \cite{shokeen2017} found there is a small
enhancement within 20 fs experimentally, so the $v_\downarrow=1.69~\rm
\AA/fs$, which is well within our expectation if we compare it with
1.44 $\rm \AA/fs$ of Ni.  In other words, the pure ballistic spin
transport contribution should be over within 20 fs. This time scale is
still too short for many prior experiments
\cite{rudolf2012,pfau2012,eschenlohr2013}.

Now with the spin enhancement time understood, we can address the spin
moment loss. Hofherr \et \cite{hofherr2017} found that the spin moment
loss in Ni is 0.52 $\mu_B$/atom, but the spin increase in the Au film
is 0.015 $\mu_B$/atom, only 2.8\%, with the 97.2\% spin loss
unaccounted for.  Given that MOKE is bulk-sensitive, such a huge
discrepancy is surprising. One possible explanation from our picture
is that the main spin loss in Ni is in the $3d$ states and is local. A
small portion of frontier $3d$ electrons, including both majority and
minority spins, enters the Au layer. The spin enhancement peak is
formed due to the arrival of minority spins; once the majority spin
leaves, the minority spin dominates and leads to the spin reversal.
{For this reason, the density of states across the Fermi
  level is crucial to the spin transport as shown recently for Gd
  \cite{jpcm17c} and in Co/Cu(001) interfaces \cite{chen2018}. More
definitive answers require a detailed calculation of density of states
at the interface between the Ni and Au layers.}
 
Finally, to quantify the amount of the spin transported into a
nonmagnetic layer, we propose a spin-valve structure. Figure
\ref{fig3}(c) shows such a structure. A ferromagnetic layer is grown
on the wedged nonmagnetic layer of length $l_{nm}$, and is pumped by a
laser pulse. One can also pump on the nonmagnetic layer. Depending on
the location that the laser beam aims at, one can systematically
control the amount of the spin current flowing into the nonmagnetic
layer by measuring the magneto-resistance in the circuit. However,
this experiment may not be easy since the electric current detection
is normally much slower than the optical stimulus, but at least this
gives some quantitative measure of how much the spin propagates into the
nonmagnetic layer. 

\section{Conclusions}

We have carried our a first-principles calculation to investigate
whether transport through the intraband transition affects the
demagnetization. We employ two systems, one bulk and one ultrathin
film. We find that in both systems the effect of transport on
demagnetization is very small, less than 1\%. The maximum velocity in
Ni is 0.4 $\rm \AA/fs$. This is much smaller than that assumed in the
SSD theory, where all the $sp$ electrons gain 10 $\rm \AA/fs$.  In
addition, the velocity oscillates strongly, so the net displacement
for the electron is very small. We should point out that it is the net
velocity gained by the electron that is related to the transport, not
the Fermi velocity, since in the crystal momentum space the velocities
should be symmetric without an external field.  The charge response is
more pronounced and also faster than the spin.  Following the latest
experimental findings \cite{shokeen2017}, we suggest the entire
demagnetization should be separated into two categories, photo-doping
and photo-excitation. In photo-excitation, the electrons are excited
to excited states and then the demagnetization starts, while in
photo-doping, the electrons transport from one material to another, so
this process depends critically on the materials in question. For
instance, whether the Fe layer is excited first or Ni layer excited
first matters to the entire demagnetization process, since they have
different Fermi energies. We trusts that our finding will motivate
further experimental and theoretical investigations.


\acknowledgments

We appreciate the helpful communications with Dr. T. T. Luu and
Dr. M. Dignam on the intraband transition \cite{luu2016,dignam2014}.
This work was solely supported by the U.S. Department of Energy under
Contract No. DE-FG02-06ER46304. Part of the work was done on Indiana
State University's high performance quantum and obsidian clusters.
The research used resources of the National Energy Research Scientific
Computing Center, which is supported by the Office of Science of the
U.S. Department of Energy under Contract No. DE-AC02-05CH11231.

$^*$gpzhang.physics@gmail.com

\newpage

\begin{figure}
\includegraphics[angle=0,width=1\columnwidth]{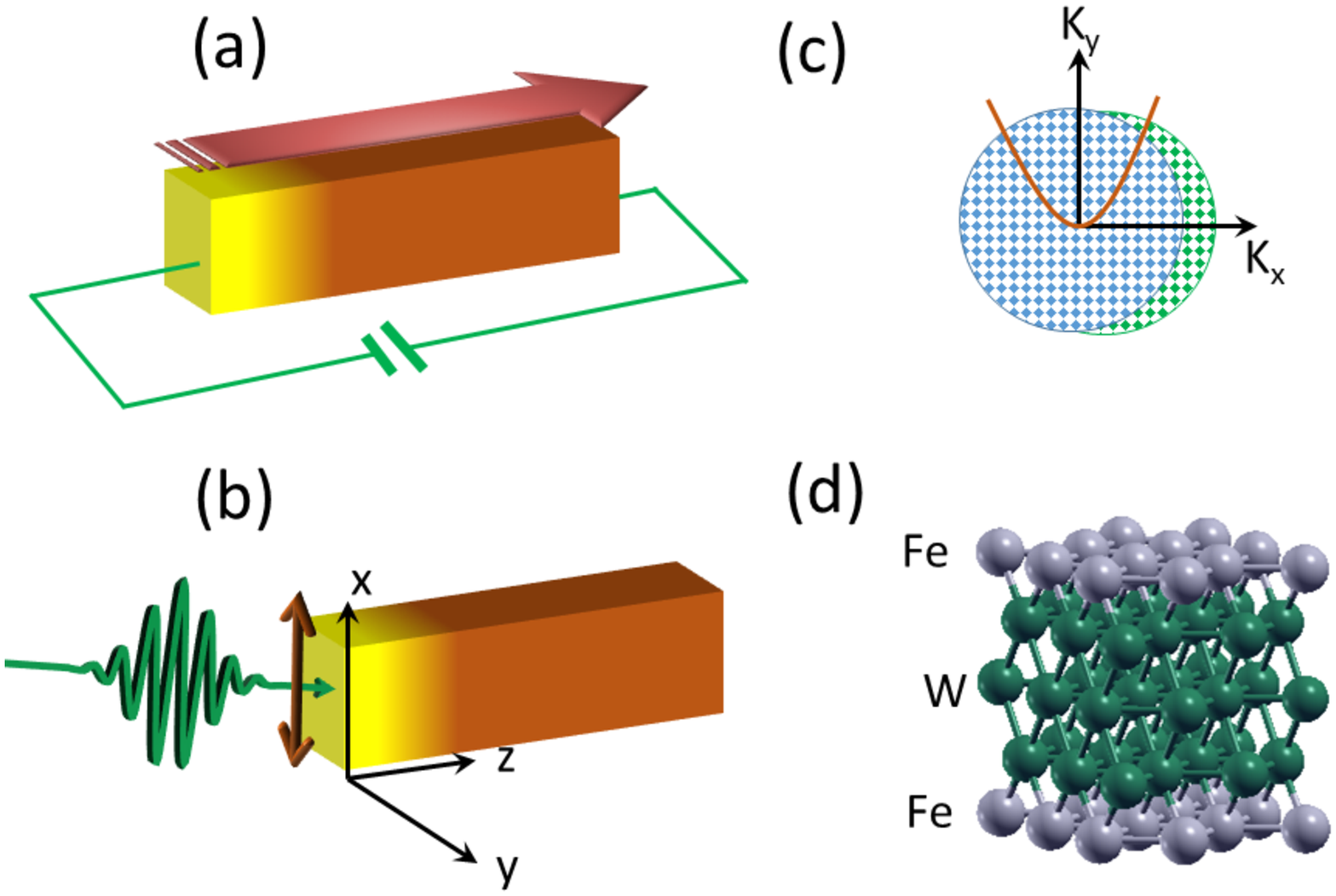}
\caption{ 
%
(a) Spin transport geometry under electric current. The bias is
  applied longitudinally, so electrons move in the opposite direction
  of the electric field. (b) Laser-induced spin transport. Here the
  laser electric field is perpendicular to the light propagation
  direction. The initial motion of the electron is vertical.
  {(c) If the interband transition is ignored, the Fermi
    sphere shifts under an external field.  However, in our
    simulation, we do not use this approach.}  (d) Supercell of one
  layer of Fe on three layers of W(110).  }
\label{fig1}
\end{figure}

\begin{figure}
\includegraphics[angle=270,width=1\columnwidth]{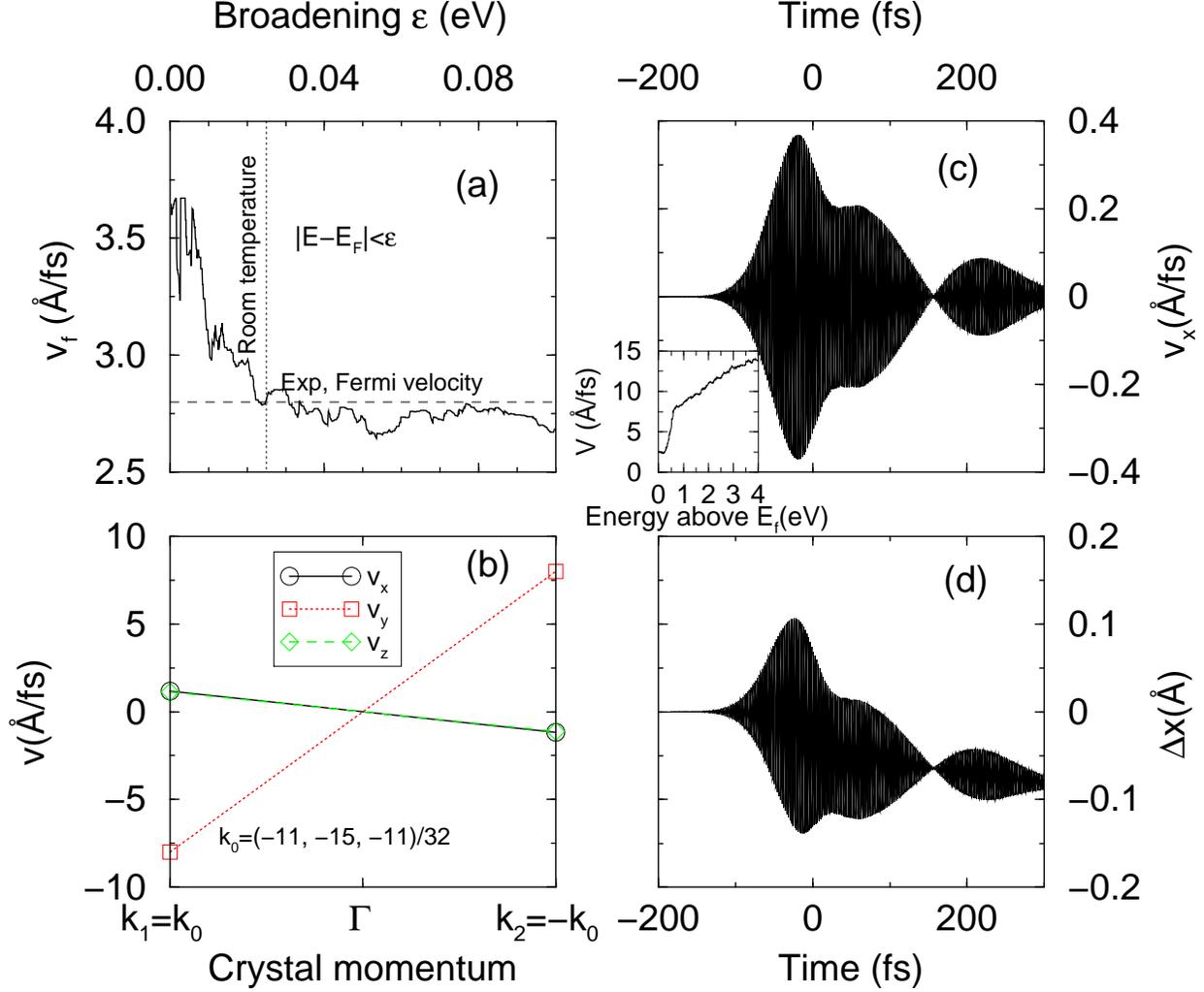}
\caption{(a) Fermi velocity as a function of energy broadening
  $\epsilon$ in fcc nickel. The horizontal dashed line is the
  experimental Fermi velocity.  (b) Electron velocity comparison
  between two \k points in opposite directions.  $\k_1=(11,15,11)/32$
  and $\k_2=(-11,-15,-11)/32$ in the unit of the reciprocal lattice
  vector $b=2\pi/a$, where $a$ is the lattice constant of fcc Ni.
  These two \k points have the largest velocity with band 70, which is
  at the Fermi level.  (c) Collective velocity along the $x$ axis upon
  laser excitation in fcc nickel.  Inset: Velocity as a function of
  energy. Here the energy is referenced to the Fermi energy.  (d)
  Collective displacement along the $x$ axis. This is calculated by
  integrating the velocity over time.
}
\label{vf}
\end{figure}

\begin{figure}
\includegraphics[angle=270,width=1\columnwidth]{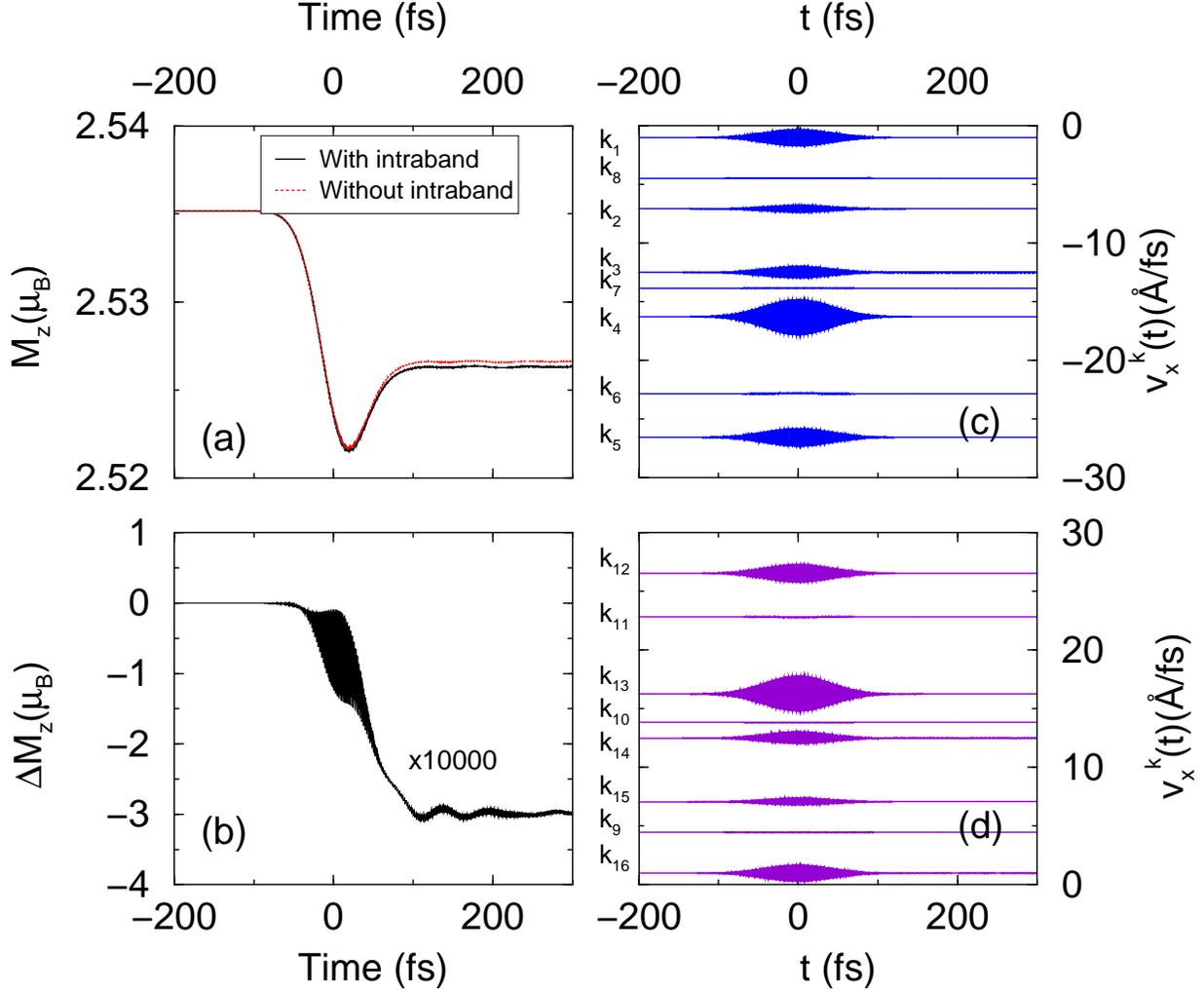}
\caption{(a) Demagnetization with and without intraband transitions
  in fcc Ni. Our laser field is applied along the $x$ axis.  The field
  amplitude is 0.03 $\rm V/\AA$ and duration is 60 fs.  Solid line:
  with intraband contribution. Dotted line: without intraband
  transitions.  Including intraband transitions increases the amount
  of demagnetization. Note that we use a simple cubic to simulate fcc
  Ni, where there are four atoms in the unit cell and the spin moment
  is four times larger than the fcc cell.  (b) Difference between two
  spin moments, where the curve is multiplied by 10000.  The
  difference is very small. (c) Crystal-momentum-dispersed velocities
  as a function of time on the first half of the $\Gamma-X$ line.
  $k_i=(i,1,1)b/32$, where $i$ runs from 1 to 15 in steps of 2.  $b$
  is the reciprocal lattice constant.  (d) Crystal-momentum-dispersed
  velocities as a function of time on the second half of the
  $\Gamma-X$ line.  $k_i=(i,1,1)b/32$, where $i$ runs from 17 to 31 in
  steps of 2.  }
\label{ni}
\end{figure}

\begin{figure}
\includegraphics[angle=270,width=1\columnwidth]{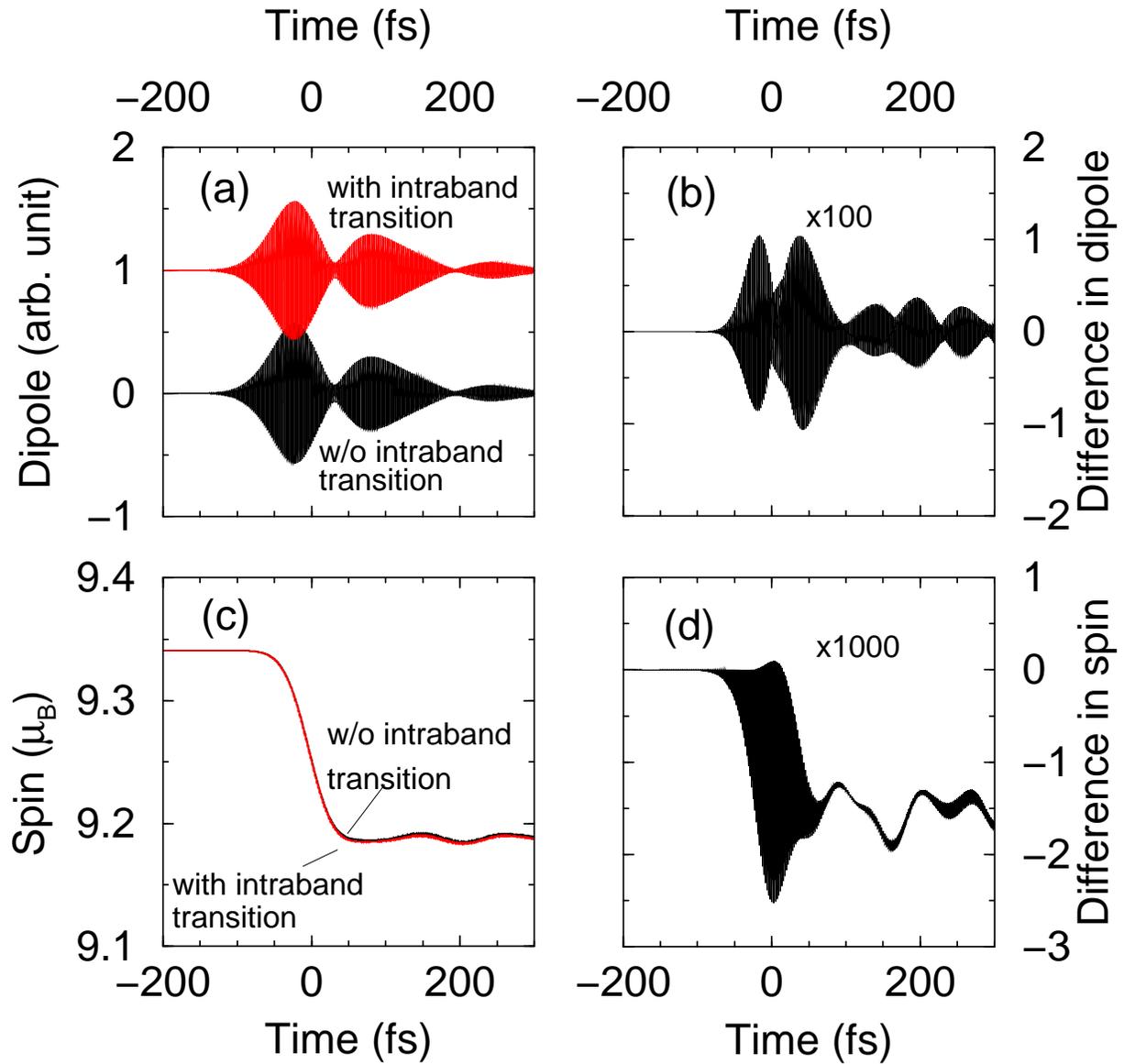}
\caption{ (a) Comparison between the dipole moments with and without
  intraband transitions in Fe/W(110). (b) Difference between the
  dipole moments, multiplied by 100. (c) Ultrafast demagnetization
  with and without intraband transitions. (d) Spin-moment difference.  }
\label{fig2}
\end{figure}

\begin{figure}
\includegraphics[angle=270,width=1\columnwidth]{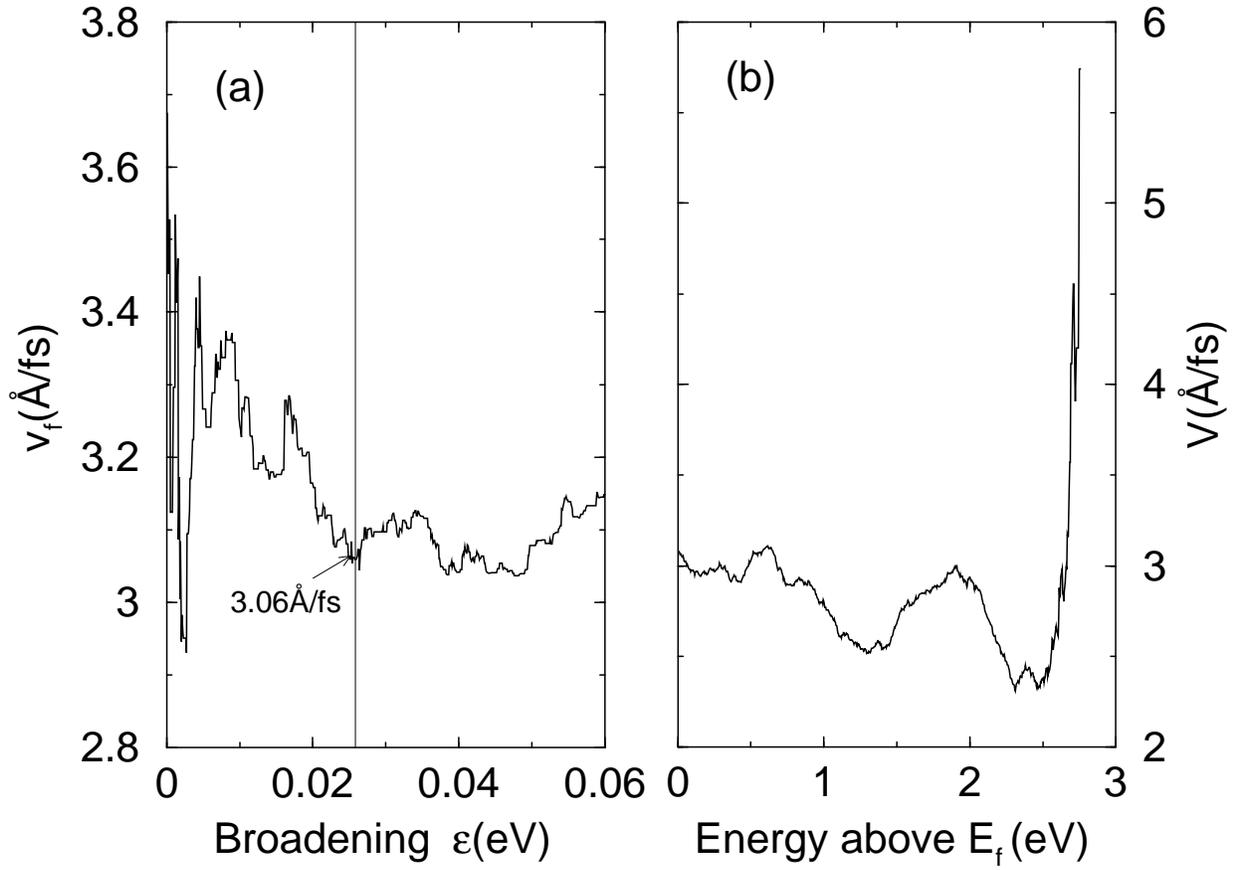}
\caption{(a) Fermi velocity as a function of the
  broadening $\epsilon$ in the Fe/W(110) thin film.  The vertical line
  denotes the room temperature, where we find the Fermi velocity is
  3.06 $\rm \AA/fs$. This is higher than that in Ni.  (b) Velocity as
  a function of the energy with respect to the Fermi energy. We
  include an energy window of 0.2 eV. We only plot the energy up to 3
  eV.  }
\label{w3fe2}
\end{figure}

\begin{figure}
\includegraphics[angle=0,width=0.7\columnwidth]{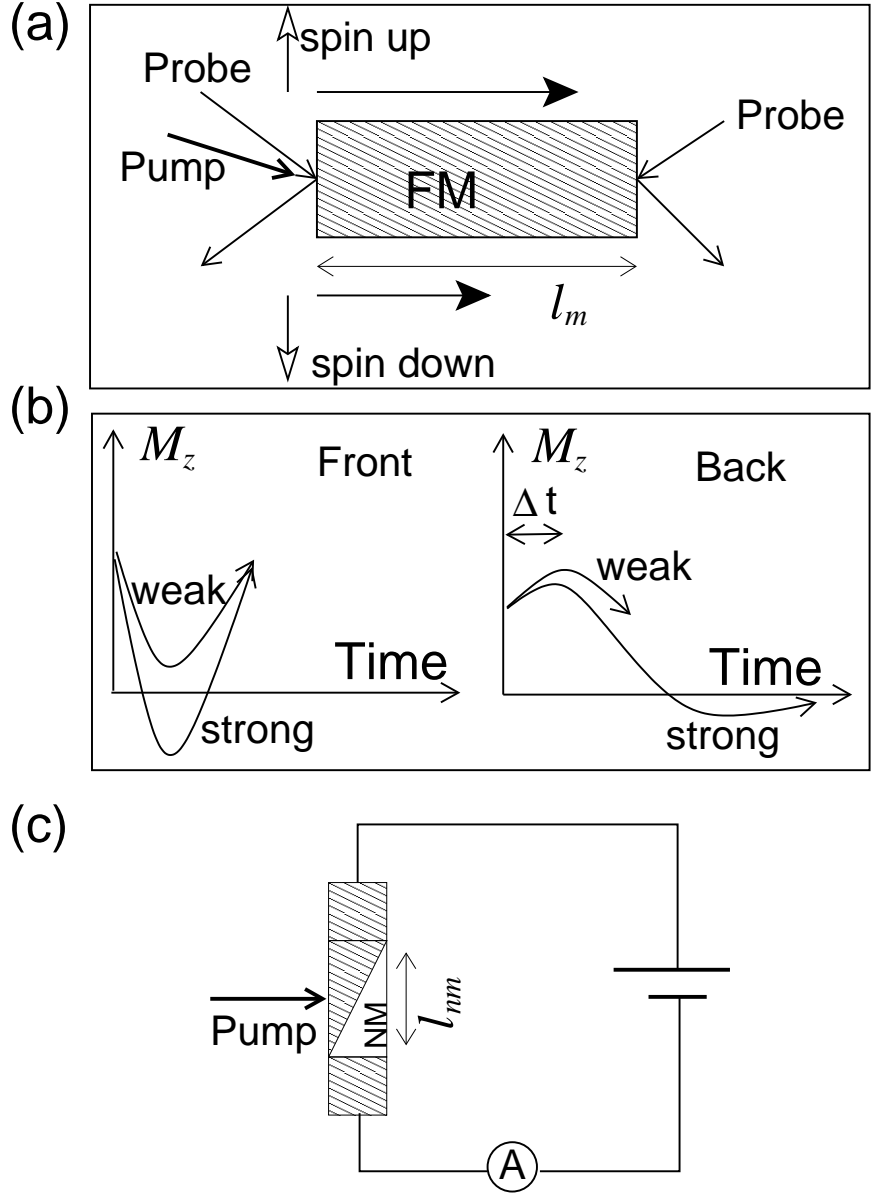}
\caption{Theoretical proposal for testing if only the pure spin
  transport contributes to the ultrafast demagnetization. (a) Geometry
  of the proposed experiment. Similar to the experimental detection
  scheme, the pump is always on the front. The detection can be either
  on the front or on the back. Majority and minority spins move at
  different velocities.  (b) Predicted effect of transport on
  demagnetization. For the front probe, the spin drops and returns to
  its original value after the minority spin departs.  For the back
  probe, one should see an enhancement within the delay between the
  majority and minority spins. (c) Proposed experiment to detect the
  spin injection into the nonmagnetic layers as a function of the
  thickness of both magnetic and nonmagnetic layers. The pump can be
  on the side of the nonmagnetic layer as well.  }
\label{fig3}
\end{figure}

{

\appendix

\section{Pure intraband transitions}

If we do not have the first term on the right side of Eq. (\ref{liu})
and only keep the diagonal terms of $\rho_\k$, then we recover the
classical Boltzmann equation.  Here the time evolution is determined
by \be \hbar \frac{\partial \rho_\k}{\partial t} +e {\bf F}(t)\cdot
\nabla_\k \rho_\k =0, \ee which is the standard first-order
homogeneous equation \cite{asmar}, \be \frac{\partial u}{\partial x} +
p(x,y) \frac{\partial u}{\partial y}=0, \ee where in general the
unknown $u$ and known $p$ are both functions of $x$ and $y$.
Mathematically, equations like this have an exact solution, which is
found by the method of characteristic curves. The key idea is that one
finds a path or curve (defined by $(x,y)$) where $u(x,y)$ is
constant. On this curve, how $y$ changes depends on how $x$
changes. Their relation is determined by the derivative of $y$ with
respect to $x$, $dy/dx=p(x,y)$.

Here is a brief
explanation.   First, let us consider a simple case where $p(x,y)=1$, so we
 have \cite{asmar}  
\be
\frac{\partial u}{\partial x}+\frac{\partial u}{\partial y}=0,
\label{od}
\ee where $u=u(x,y)$ is the unknown function. According to Asamar
\cite{asmar}, if $f$ is any differentiable function of a single
variable, then \be u(x,y)=f(x-y) \label{eq1}\ee is a solution of
Eq. (\ref{od}). We can verify this by using the chain rule, where we get
that \be \frac{\partial u}{\partial x}=f'(x-y); \hspace{1cm}
\frac{\partial u}{\partial y}=-f'(x-y). \label{eq2} \ee We see that
Eq. (\ref{od}) holds with this solution. Note that the functions are
constant on any line $(x-y=c)$ due to the form of the solution
$u(x,y)=f(x-y)$.  Here $c$ is constant. The actual form of $u(x,y)$ is
determined by the initial condition. One example in physics is the
Fermi function $ f(E)=\frac{1}{1+\exp(E-E_f)}, $ where $E$ is the band
energy and $E_f$ is the Fermi energy.

If $p(x,y)=2$, then the variable in Eqs. (\ref{eq1}) and (\ref{eq2})
is $(2x-y)$. In other words, on the characteristic line $x$ must move
two units for every one unit along the $y$ direction. For all the
other cases, one can derive a similar relation.

Next, we consider a generic case with nonconstant $p(x,y)$.  Here the
above line becomes a curve where $x$ and $y$ change according to the
constraint $dy/dx=p(x,y)$.  In other words, the rate of change of $y$
with respect to $x$ is just $dy/dx=p(x,y)$. On the curves $(x,y)$ with
the constraint $dy/dx=p(x,y)$, $u$ is constant. If we suppose the
solution of $dy/dx=p(x,y)$ is $\phi(x,y)$, as far as $\phi(x,y)$ is
constant on the characteristic curve, we have a solution
$u(x,y)=f(\phi(x,y))$, where the functional of $f$ can be an arbitrary
function. As discussed above, the actual form is fixed by the initial
condition, in our case, a Fermi function, so $\rho_\k$ does not change
its shape, regardless of what the external field looks like.  When we
apply $dy/dx=p(x,y)$ to our problem, $p$ is $e{\bf F}(t)/\hbar$, so we
get \be \frac{\partial \k}{\partial t}=\frac{e {\bf F}(t)}{\hbar}=-
\frac{e}{\hbar} \frac{ {\bf A}(t)}{\partial t}, \label{vector} \ee
which, after integration, leads to our familiar form of \be {\bf k}+ e
{\bf A}(t)/\hbar = {\bf k}_0.  \ee ${\bf A}(t)$ in Eq. (\ref{vector})
is the vector potential of the laser field ${\bf F}(t)$.  If we
relabel the original \k in $\rho_{\k}$ by $\k_0$, our solution is
$\rho_\k=\rho_\k(f({\bf k}+e {\bf A}(t)/\hbar))$.  Here, in absence of
interband transitions, $\rho_\k$ is always a Fermi distribution
function.  This solution is exact mathematically, independent of the
form of ${\bf A}(t)$, constant or oscillatory.  Physically, in the
absence of interband transitions, such transitions are equivalent to
shifting the electron Fermi surface in the reciprocal space along the
external field direction. Figure \ref{fig1}(c) illustrates such a
situation. The amount of shift is determined by the laser vector
potential. With the presence of the interband transitions, such
shifting no longer works, so one has to use Eq. (\ref{liu}), which is
exactly what we do here. }

%


\end{document}